\documentclass{ws-p8-50x6-00}

\begin{document}

\title{Elastic Form Factors in Point-form Approach}

\author{Marco Radici}

\address{Dipartimento di Fisica Nucleare e Teorica, Universit\`a di Pavia \\
and Istituto Nazionale di Fisica Nucleare, Sezione di Pavia, I-27100 Pavia, 
Italy}


\maketitle

\abstracts{Using the point-form approach to relativistic quantum 
mechanics, a covariant framework is presented for the calculation 
of proton and neutron electromagnetic form factors. Results for
charge radii, magnetic moments, and electric as well as magnetic 
form factors are produced using the wave functions obtained in the 
so-called Goldstone-boson-exchange constituent quark model. All the 
predictions are found in remarkable agreement with existing 
experimental data.}

\section{Introduction}
\label{sec:intro}

The key issue for a QCD-motivated description of low-energy strong 
interactions is the identification of the effective degrees of 
freedom that describe the hadron phenomena. By considering the features
of the baryon spectrum, the well-known spontaneous breaking of chiral
symmetry suggests that below a certain energy threshold the effective
``particles'' to be considered should be the constituent quarks, whose
dynamical mass is related to the $\langle q \bar q \rangle$ condensate,
and the Goldstone bosons arising from the dynamical symmetry breaking
and coupling directly to the constituent quarks.

The Constituent Quark Model (CQM) based on the assumption that the
quark-quark interaction is mediated by the Goldstone Boson Exchange
(GBE)~\cite{riska}, is capable of giving a consistent description of the 
low-energy spectra of light and strange baryons~\cite{graz}. It assumes 
a linear confinement, as suggested by lattice QCD, and retains only the 
spin-spin component of the pseudoscalar GBE hyperfine interaction. 

However, the CQM should also provide a comprehensive description of
other hadron phenomena. As a first test of the model wave functions, 
here we consider the properties of the nucleon that can be explored with 
an electromagnetic probe: charge radii, magnetic moments and elastic 
form factors. In order to get a reliable test, the uncertainties
introduced in the calculation of the scattering amplitude need to be
minimized. To this aim, a proper treatment of the relativistic
kinematics is required. The suggested value of the mass parameter for
the constituent quark indicates that traditional nonrelativistic
expansions in $p/m$ are not justified~\cite{graz1}. Moreover, as we
will see later, the effects due to the Lorentz boost of the three-quark
system are crucial~\cite{tutti}. 

Among the various possibilities of setting up a relativistic quantum
theory~\cite{dirac}, here we choose the point-form formulation. It is 
characterized by several distinctive features. All the dynamics is 
contained in the four-momentum operators, which commute among themselves 
and can be simultaneously diagonalized. The generators of the Lorentz 
boosts contain no interactions and, therefore, are purely kinematic; the 
theory is thus manifestly covariant. In practice, the point-form 
formulation allows for a proper treatment of the Lorentz boosts of the 
three-quark wave functions and for an accurate calculation of the matrix 
elements of the electromagnetic current operator~\cite{klink}. Due to 
the fact that the GBE CQM uses a relativistic kinetic energy operator, 
the full Hamiltonian leads to a mass operator fulfilling all the 
necessary commutation relations of the Poincar\'e group~\cite{keistpoly}, 
even if the quark-quark interaction consists of a phenomenological 
confinement and of an instantaneous GBE potential. 


\section{The point-form approach to the scattering amplitude}
\label{sec:formulae}

The starting point are the solutions of the eigenvalue problem for the
Hamiltonian 
\begin{equation}
\label{eq:hamil}
H=\sum\limits_{i=1}^3\sqrt{\vec{k}_i^2+m_i^2}+\sum\limits_{i<j=1}^3
\left[V^{\rm conf}(i,j)+V^{\rm GBE}(i,j)\right],
\end{equation}
where $m_i$ are the masses and $\vec{k}_i$ the three-momenta of the 
constituent quarks in the nucleon rest frame, respectively. The
confinement interaction $V^{\rm conf}$ and the spin-spin component of
the pseudoscalar GBE hyperfine interaction $V^{\rm GBE}$ produce a
low-energy baryon spectrum in remarkable agreement with experimental
results~\cite{graz}. 

The three-quark wave functions refer to constituent quarks in the
nucleon rest frame, i.e. with total momentum $\vec{P}=0$. Therefore,
they can be interpreted also as eigenstates of the mass operator
including interactions~\cite{klink}
\begin{equation}
\label{eq:mass}
M=\sqrt{P^\mu P_\mu}=M_{\rm free}+M_{\rm int},
\end{equation}
where $P^\mu=P^\mu_{\rm free}+P^\mu_{\rm int}$ is the four-momentum
operator with interactions inserted according to the Bakamjian-Thomas
(BT) construction~\cite{BT} in point form. The most general 
representation of such eigenstates is defined in the product space 
${\mathcal H}_1\otimes{\mathcal H}_2\otimes{\mathcal H}_3$ of 
one-particle spin-$1\over 2$, positive-mass, positive-energy 
representations of the Poincar\'e group
\begin{equation}
\label{eq:prodstate}
\left|\psi\right\rangle \equiv 
\left|p_1,\lambda_1\right\rangle\otimes
\left|p_2,\lambda_2\right\rangle\otimes
\left|p_3,\lambda_3\right\rangle,
\end{equation}
where $p_i$ are the individual quark four-momenta and $\lambda_i$ the
$z$-projections of their spins. A general Lorentz transformation
$U_\Lambda$ on $|\psi\rangle$ produces three different Wigner rotations.
It is more convenient to first introduce so-called velocity
states~\cite{klink} by applying a particular Lorentz boost $U_{B(v)}$ to 
the center-of-momentum states, which are defined analogously to
Eq. (\ref{eq:prodstate}) but fulfil the constraint 
$\vec{P}=\vec{k_1}+\vec{k_2}+\vec{k_3}=0$,
\begin{equation}
\label{eq:vstate}
\left|v;\vec{k}_1,\vec{k}_2,\vec{k}_3;\mu_1,\mu_2,\mu_3\right\rangle =
U_{B(v)}\left|k_1,k_2,k_3;\mu_1,\mu_2,\mu_3\right\rangle 
= \prod\limits_{i=1}^3D^{1/2}_{\lambda_i\mu_i}[R_W(k_i,B(v))]
\left|\psi\right\rangle 
\end{equation}
and $p_i=B(v)k_i$. Under a general Lorentz transformation $U_\Lambda$
and for canonical boosts, the velocity state (\ref{eq:vstate})
transforms as
\begin{eqnarray}
\label{eq:boostv}
&&U_\Lambda\left|v;\vec{k}_1,\vec{k}_2,\vec{k}_3;\mu_1,\mu_2,\mu_3
\right\rangle= \nonumber\\
&=&\prod\limits_{i=1}^3D^{1/2}_{\mu'_i\mu_i}(R_W) 
\left|\Lambda v;R_W\vec{k}_1,R_W\vec{k}_2,R_W\vec{k}_3;\mu'_1,\mu'_2,
\mu'_3\right\rangle;
\end{eqnarray}
the Wigner rotations $R_W$ are now all the same and the spins can thus 
be coupled together to a total spin as in nonrelativistic theory.

After setting up the proper Lorentz boosts of the nucleon state, the
next task is to calculate the matrix element of the electromagnetic
current operator. Following the formalism of Ref.~\cite{klink} , it is 
possible to show that
\begin{eqnarray}
\langle\psi'(P',\sigma')| J^{\mu}(0) |\psi(P,\sigma)\rangle &= 
&\displaystyle{\int} d^4Q \, B(Q)^{\mu}_b \, \langle\psi'(P',\sigma')| 
J^b(Q) |\psi(P,\sigma)\rangle \nonumber \\
& &\hspace{-3.5cm} = \displaystyle{\int} d^4Q \, \delta^4(P'-P-Q) 
\displaystyle{\sum_{b,r',r}} \left( P\frac{1}{2} \sigma \, ; \, Qbr'r 
\vert P'\frac{1}{2} \sigma' \right) \times F^b_{r' r}(Q^2) ,
\label{eq:matrel}
\end{eqnarray}
where $b=0,1,2$ for $Q^2<0$ and $r,r'=\pm \frac{1}{2}, |r-r'|=0,\pm 1$.
$J^b$ is an irreducible tensor of the Poincar\'e group, ensuring that
$J^{\mu}$ is covariant and conserved. Moreover, the Wigner-Eckhart
theorem allows for the replacement of the matrix element of $J^b$ by a 
linear combination of products of Clebsh-Gordan coefficients of the
Poincar\'e group (linking a time-like $P$ and a space-like $Q$ to a
time-like $P'$) and of reduced matrix elements $F^b_{r'r}$, which can be
identified with the invariant form factors. By choosing the convenient
Breit frame with the 3-momentum along the $\hat z$ axis, the initial and
final velocities are given by 
$m_N v_i=P_{\rm st}=(\sqrt{m_N^2+(Q/2)^2},0,0,-Q/2)$ and
$m_N v_f=P'_{\rm st}=(\sqrt{m_N^2+(Q/2)^2},0,0,Q/2)$, 
respectively, where $m_N$ denotes the nucleon mass. 

Assuming the Impulse Approximation (IA), i.e. assuming that the virtual 
photon hits the quark labelled "1" leaving the "2","3" as spectators, 
the invariant form factors of Eq. (\ref{eq:matrel})
become~\cite{klink,tutti}
\begin{eqnarray}
\begin{array}{l}
F^b_{r'r} = 3 \displaystyle{\int} d\vec k_2 d\vec k_3 d\vec k'_2 
d\vec k'_3 \, \psi^*_{r'}(k'_2 \mu'_2, k'_3 \mu'_3) \, 
\psi_r (k_2 \mu_2, k_3 \mu_3) \\
\qquad \begin{array}{l}
\delta^3 [k'_2 - B^{-1}(v_f) B(v_i)k_2]
\, \delta^3 [k'_3 -B^{-1}(v_f) B(v_i)k_3] \\[5pt]
\begin{array}{l}
D^{\frac{1}{2}}_{\mu'_2 \mu_2} [R_W (k_2, B^{-1}(v_f) B(v_i))] \, 
D^{\frac{1}{2}}_{\mu'_3 \mu_3} [R_W (k_3, B^{-1}(v_f) B(v_i))]
\\[5pt]
\hspace{-1cm} \begin{array}{l} 
D^{\frac{1}{2}\, *}_{\lambda'_1 \mu'_1} [R_W (k'_1, B(v_f))] \, 
\langle p'_1 \lambda'_1 | (\gamma^b f_1 + 
i \displaystyle{\frac{\sigma^{b\nu} (p'_{1\nu}-p_{1\nu})}{2m_1}} f_2) 
| p_1 \lambda_1\rangle \, D^{\frac{1}{2}}_{\lambda_1 \mu_1} 
[R_W (k_1, B(v_i))] 
\end{array}\end{array} \end{array} \end{array} ,
\label{eq:invff} 
\end{eqnarray}
where the standard form for the relativistic single-particle current 
matrix element for quark label "1" has been used, that contains also 
the quark form factors $f_1(Q^2), f_2(Q^2)$. However, the following results 
have been obtained assuming point-like constituent quarks, i.e. $f_1(Q^2)=1$ 
and $f_2(Q^2)=0$.

From Eq.~(\ref{eq:invff}) one obtains the nucleon Sachs form factors 
through~\cite{klink}
\begin{equation}
\label{eq:sachs}
\left.\begin{array}{l}
\displaystyle F^{b=0}_{r'r}(Q^2)=G_E(Q^2)\delta_{r'r}\\
\displaystyle F^{b=1}_{r'r}(Q^2)=(r'-r) \frac{Q}{m_N}G_M(Q^2)
\delta_{r',r\pm 1}
\end{array}\right.\qquad r,r'=\pm\frac{1}{2}.
\end{equation}
Note that only the $b=0$ and $b=1$ components of $F^b_{r'r}$
are needed for the electric and magnetic form factors, respectively.
There is no new information in $F^{b=2}_{r'r}$. Therefore, the general
result is recovered that for a
spin-$\frac{1}{2}$ state only two independent form factors are needed.
The structure of the Poincar\'e group allows for an easy generalization
of this result in the point-form framework, where the matrix element of
Eq.~(\ref{eq:invff}) for a spin-$j$ state is parametrized in terms of
$2j+1$ independent reduced matrix elements~\cite{klink}. 

\section{Results}
\label{sec:res}

\begin{figure}[t]
\epsfxsize=8cm
$
\begin{array}{lr}
\hspace*{-1.5cm}\epsfbox{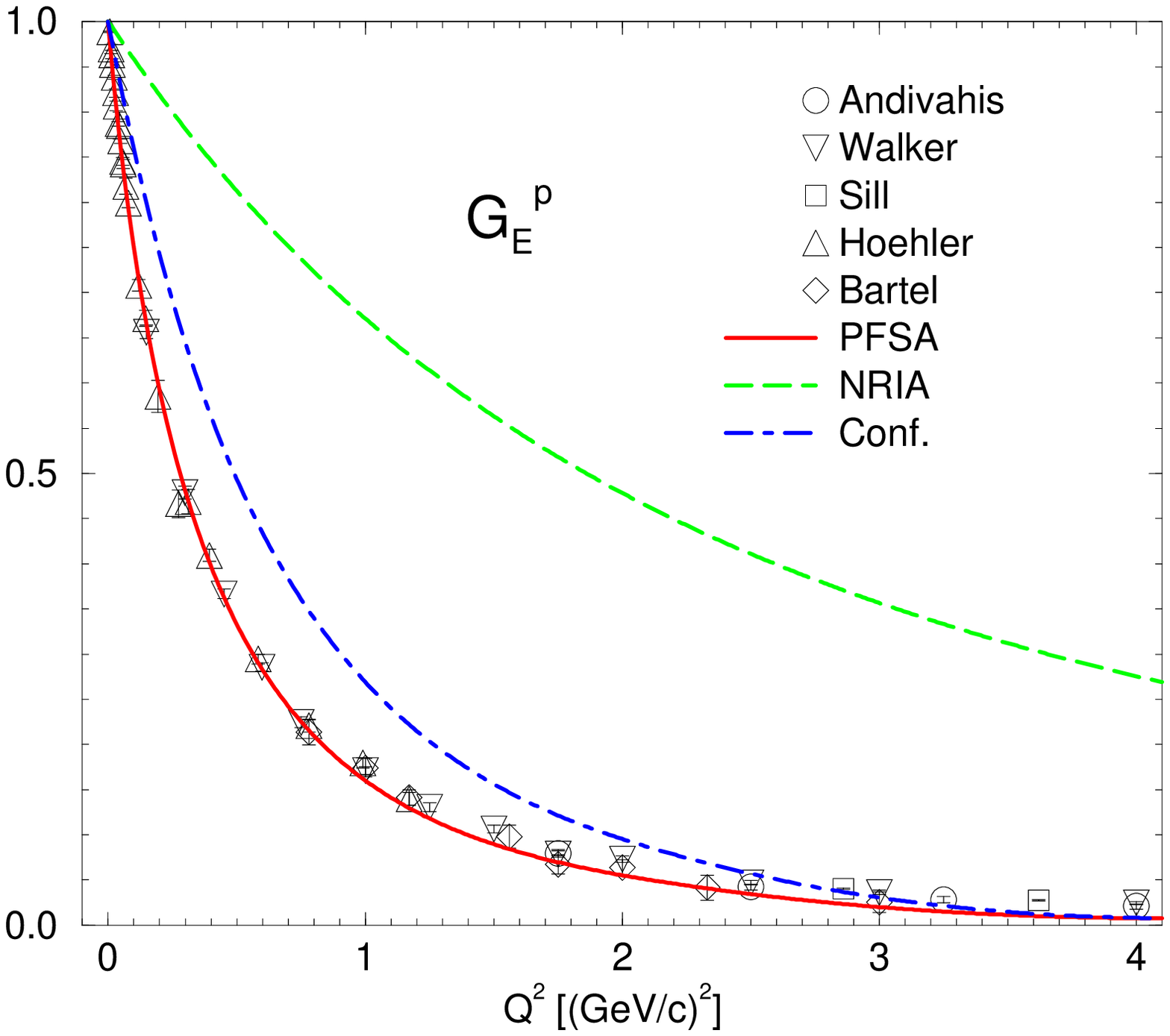} &
\hspace*{-1.3cm}\epsfbox{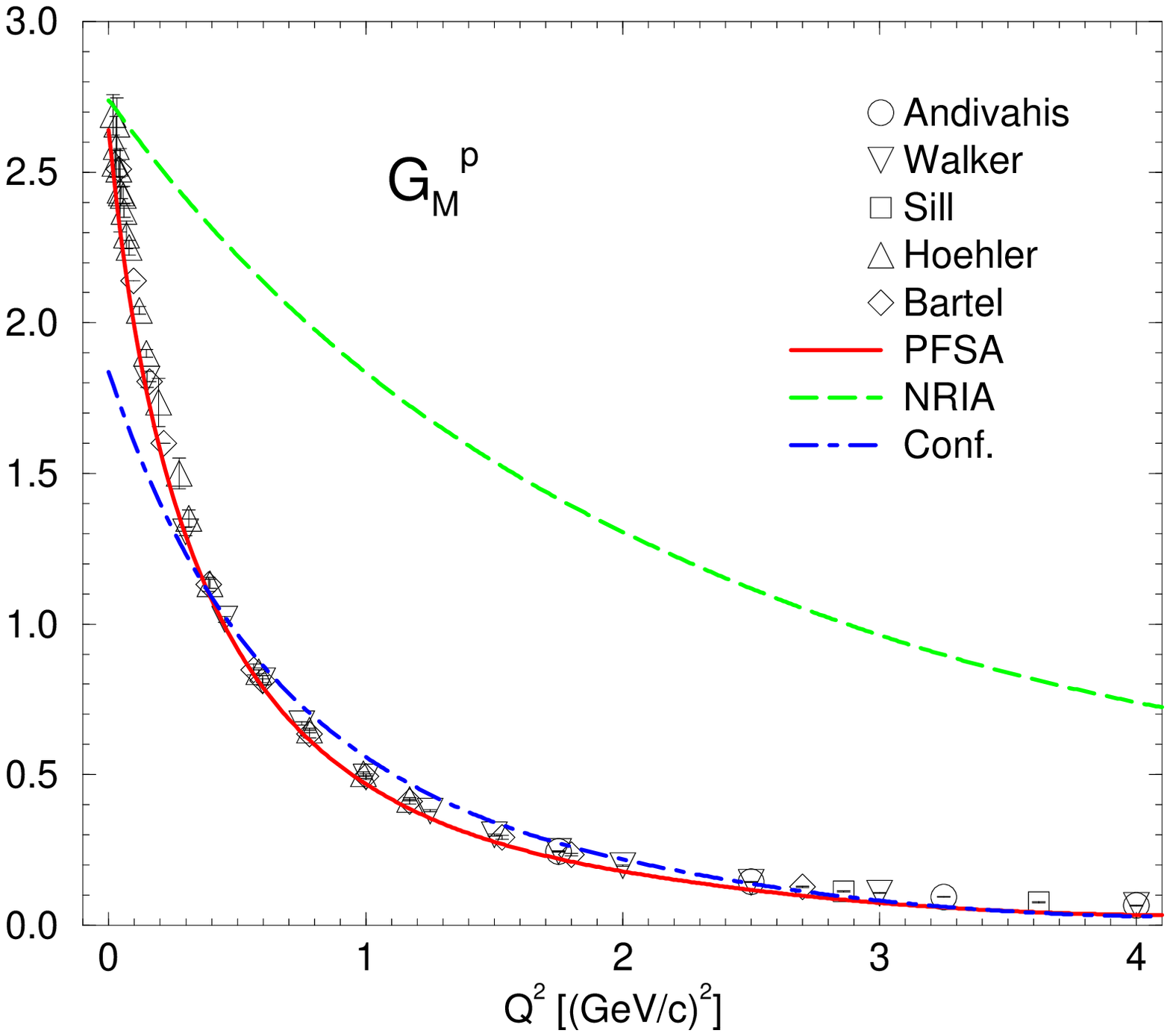}\\[-0.9cm]
\hspace*{-1.5cm}\epsfbox{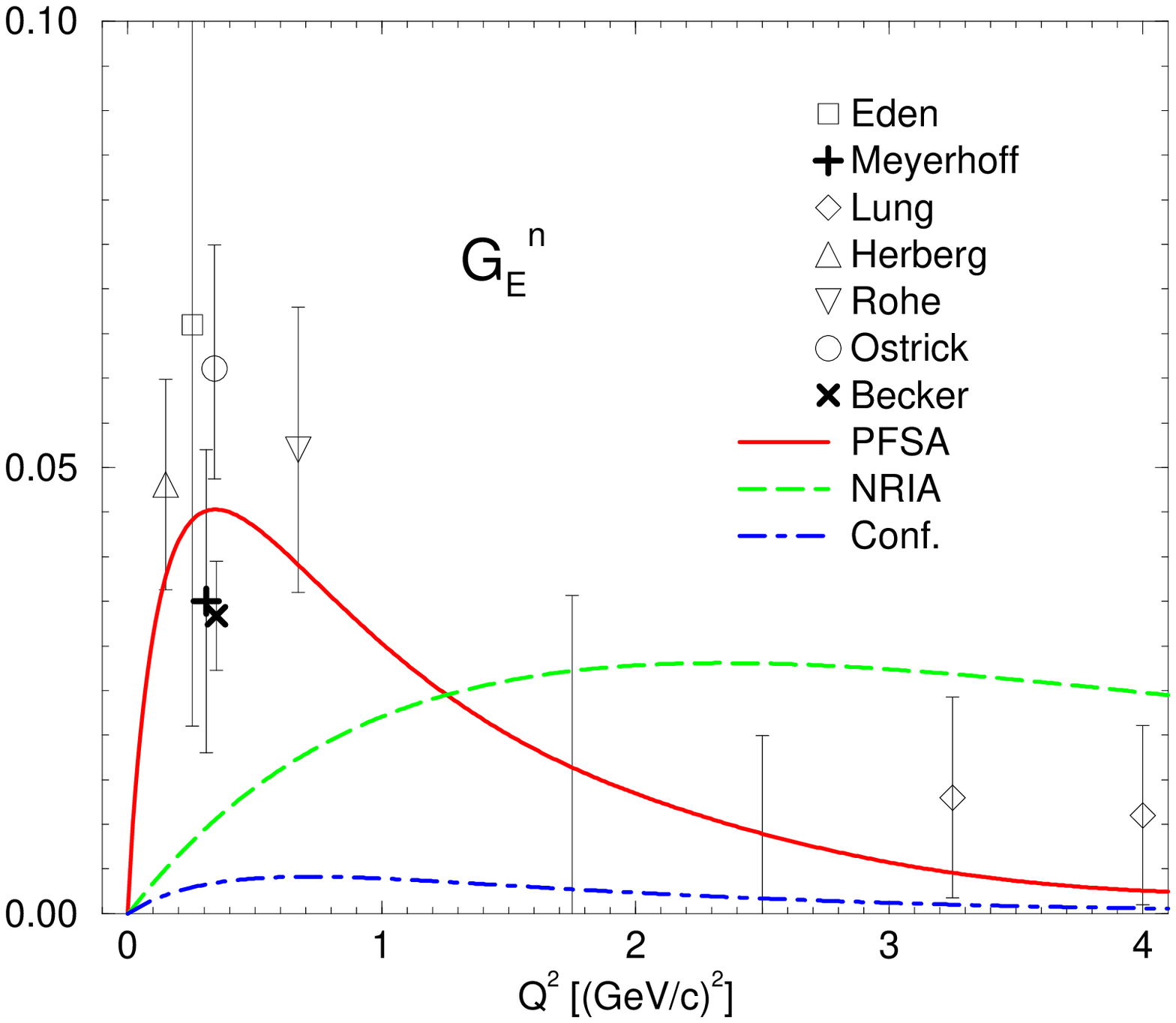} &
\hspace*{-1.3cm}\epsfbox{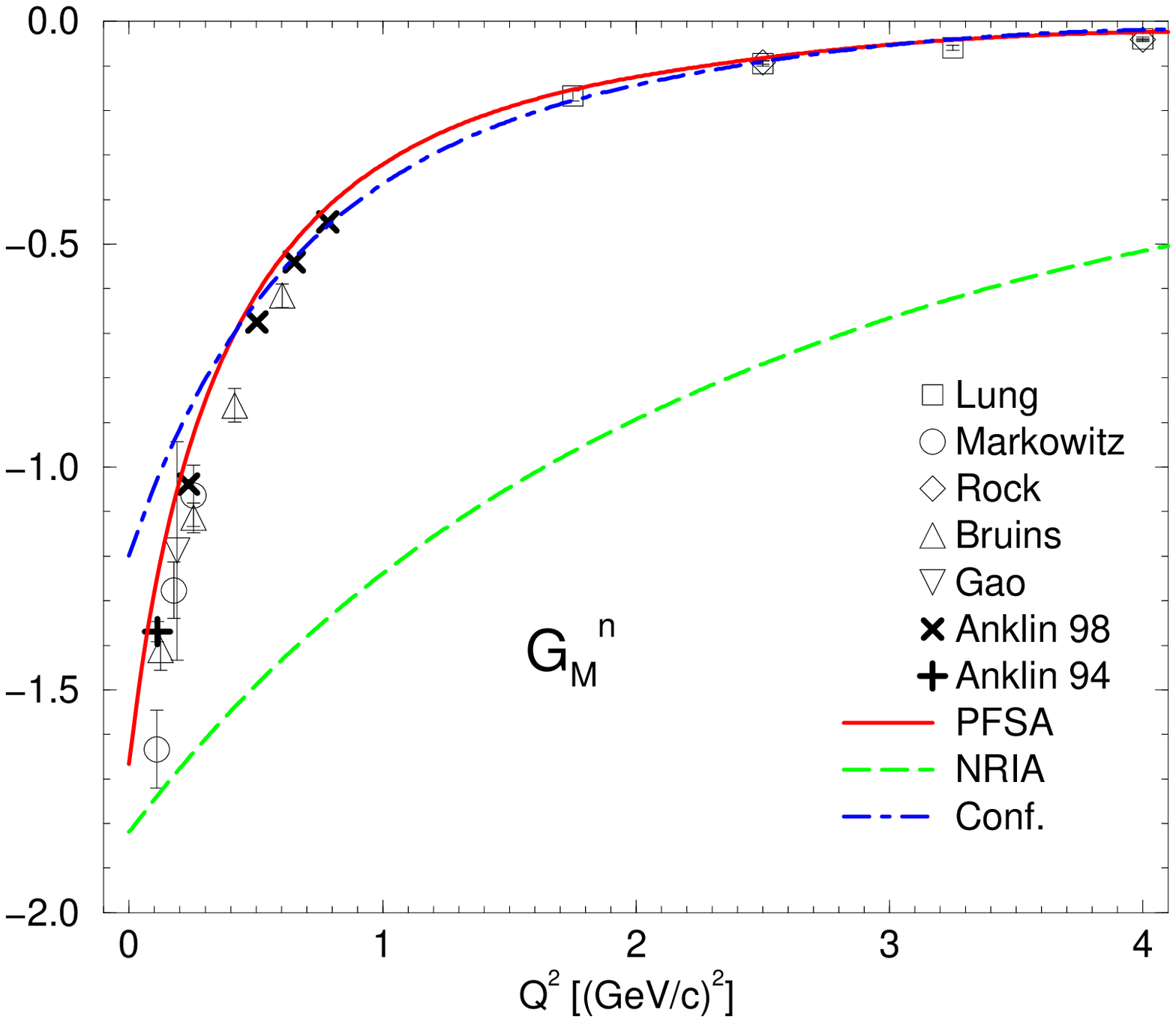}
\end{array}
$
\vspace{-1cm}
\caption{Proton (upper) and neutron (lower) electric (left) and 
magnetic (right) form factors as predicted by the GBE CQM~\protect\cite{graz} 
in IA (PFSA, solid lines). A comparison is given to the nonrelativistic results 
(NRIA, dashed) and the case with the confinement interaction only 
(dashed-dotted)~\protect\cite{tutti}. The experimental data are from
Ref.~\protect\cite{expff}. \label{fig:ff}}
\vspace{-.8cm}
\end{figure}

The predictions of the GBE CQM~\cite{graz} for the nucleon 
electromagnetic form factors are shown by the solid line (PFSA) in 
Fig.~\ref{fig:ff}. Their properties at zero momentum transfer are 
reflected by the charge radii and magnetic moments given in 
Table~\ref{tab:crmm}. The results were calculated along the lines 
explained in the previous Section. The input consists only of the 
proton and neutron three-quark wave functions as produced by the 
solution of the eigenvalue problem for the Hamiltonian (\ref{eq:hamil}).

A very good description of both the proton and neutron electromagnetic 
structure is achieved, particularly at low momentum transfer. The dashed
line represents the nonrelativistic IA result (NRIA), i.e. with the 
standard nonrelativistic form of the current operator and no Lorentz 
boosts applied to the nucleon wave functions. Evidently, relativity 
plays a major role here. The importance of proper Lorentz boosts can be 
guessed by the $\delta$ functions acting on the spectator quarks in
Eq.~(\ref{eq:invff}): $k'_i = B^{-1}(v_f) B(v_i)k_i \quad i=2,3$. At low
momentum transfer, this constraint becomes $k'_i \simeq k_i -
\frac{\omega_i}{m_N} Q$, with $\omega_i$ the energy of the $i$-th
spectator quark, and it has to be compared with the nonrelativistic
result $k'_i = k_i - \frac{1}{3} Q$~\cite{graz1}. The expectation value 
of $\omega_i$ for the employed nucleon wave functions is 
$\langle \omega_i \rangle =768$ MeV. Therefore, $\frac{\omega_i}{m_N} 
> \frac{1}{3}$, because even for low $Q^2$ the individual quark 
momenta are large. This result turns into a faster fall off of 
$G_E(Q^2)$ and a consequent bigger charge radius. 

\begin{table}[t]
\vspace{-1cm}
\caption{Proton and neutron charge radii and magnetic moments as 
predicted by the GBE CQM~\protect\cite{graz} in IA (PFSA). A comparison is 
given also to the nonrelativistic results (NRIA) and the case with the 
confinement interaction only~\protect\cite{tutti}. \label{tab:crmm}}
\begin{center}
\begin{tabular*}{\textwidth}
{l@{\qquad}r@{\qquad}r@{\qquad}r@{\extracolsep\fill}l}
\hline
                  &  PFSA    &  NRIA    &  Conf.   & Experimental\\ 
\hline
$r^2_p$ [fm$^2$]  & $ 0.75$ & $ 0.10$ & $ 0.37$ & $0.774(27)$
                                              \cite{rp-rosen},
					          $0.780(25)$
					      \cite{rp-meln}\\
$r^2_n$ [fm$^2$]  & $-0.12$ & $-0.01$ & $-0.01$ & $-0.113(7)$
                                               \cite{rn-kop}\\
$\mu_p$ [n.m.]    & $ 2.64$ & $ 2.74$ & $ 1.84$ & $2.792847337(29) $
                                                \cite{mu}\\
$\mu_n$ [n.m.]    & $-1.67$ & $-1.82$ & $-1.20$ & $-1.91304270(5)$
                                                \cite{mu}\\
\hline
\end{tabular*}
\end{center}
\vspace{-.5cm}
\end{table}

\begin{figure}[t]
\epsfxsize=9cm
\epsfbox{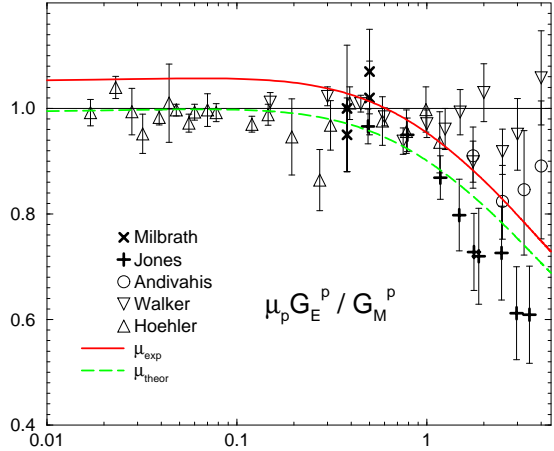}
\vspace{-1cm} 
\caption{The $\mu G_E/G_M$ ratio for the proton. Upper curve obtained
with experimental $\mu=2.79$, lower curve with theoretical $\mu=2.64$,
as given in Table~\protect\ref{tab:crmm}. Experimental data from
Ref.~\protect\cite{halla}. \label{fig:cebaf}}
\vspace{-1cm}
\end{figure}

Noteworthy, the results here presented are obtained with point-like 
constituent quarks. At least for the range of momentum transfers 
considered in Fig.~\ref{fig:ff}, there is no need to introduce 
constituent quark form factors (or any other phenomenological 
parameters beyond the CQM)~\cite{tutti}, at variance with other previous 
relativistic studies~\cite{roma,indiana,coester}. It is only with regard 
to the magnetic moments that there remains a small difference between 
the theoretical predictions and the experimental data. All this turns 
out in a underestimation of the dipole profile of $G_M(Q^2)$ and, 
consequently, in an overestimation of the recent CEBAF data for 
$\mu G_E/G_M$ for the proton~\cite{halla}, as shown by the upper curve 
in Fig.~\ref{fig:cebaf}. The lower curve is produced by the same 
calculation but replacing the correct experimental value of $\mu$
by the theoretical value $\mu=2.64$ consistent with the calculation of
$G_E,G_M$ (see Table~\ref{tab:crmm}). It indicates that there is no
violation of charge conservation (as it could wrongly appear from the
upper curve for $Q^2 \rightarrow 0$), rather it gives a "graphical"
representation of the present discrepancy of calculations with
experimental data for $G_M(Q^2)$. 

In order to get an idea of the role of the GBE hyperfine interaction in 
the electromagnetic form factors, calculations involving only the
confinement potential have been performed~\cite{tutti} and are shown by 
the dot-dashed line in Fig.~\ref{fig:ff}. All observables seem already 
in reasonable agreement, as well as charge radii and magnetic momenta. 
Only the charge structure of the neutron would remain much too small due 
to the absence of a mixed-symmetry component in the wave function. 
Though this breaking of SU(6) symmetry brought about by the hyperfine
interaction is very small~\cite{graz1}, it plays a crucial role in 
reproducing the charge radius and $G_E(Q^2)$ for the
neutron~\cite{tutti}. 

\vspace{-.3cm}
\section{Conclusions}
\label{sec:end}

The theoretical predictions for elastic nucleon form factors obtained in 
the point-form approach and using the GBE interaction~\cite{graz}, are 
found to be in remarkably good agreement with all experimental data 
(charge radii, magnetic moments, electric and magnetic form factors) 
both for the proton and the neutron. No further ingredients beyond the 
quark model wave functions (such as constituent quark form factors etc.) 
have been employed. Only relativistic boost effects are properly 
included in point-form relativistic quantum mechanics~\cite{tutti}. 

The matrix elements of the electromagnetic current operator have been
computed in IA. Current conservation in Breit frame requires the $b=3$ 
component of the current operator to vanish identically. This is not 
true in IA~\cite{klink,tutti}. However, Eq.~(\ref{eq:sachs}) shows that 
physical observables are linked to components that are not affected by 
this approximation. Therefore, the effect of many-body current operators
(needed to restore current conservation) should hopefully be small, and
could be responsible for the small discrepancy observed in the values of
magnetic momenta. 

The present results have been obtained also assuming point-like
constituent quarks. This suggests that these quasi-particles should be
anyway small. It will be very interesting to explore this possibility, 
together with a necessary exploration of nucleon electromagnetic 
transitions to other resonances in a consistent fashion.

\vspace{-.7cm}
\section*{Acknowledgments}
Numerous discussions with S.~Boffi, L.Ya.~Glozman, W.~Klink, W.~Plessas
and R.F.~Wagenbrunn are greatly acknowledged.

\end{document}